\documentclass[aps,prl,reprint,superscriptaddress,nobibnotes,longbibliography,floatfix]{revtex4-2}
\usepackage{amsmath,amssymb,graphicx,booktabs}
\usepackage{newtxtext,newtxmath,bm}
\usepackage[hidelinks]{hyperref}
\newcommand{\Tr}{\operatorname{Tr}}
\newcommand{\dd}{\mathrm d}
\newcommand{\ii}{\mathrm i}
\newcommand{\cC}{\mathcal C}
\newcommand{\e}{\mathrm e}
\newcommand{\MF}{\mathrm{MF}}
\newcommand{\xc}{\mathrm{xc}}

\hypersetup{pdftitle={An Interacting-Response Constraint for Photon Density Functionals},pdfauthor={Xiaoyu Zhang}}
\begin{document}
\title{An Interacting-Response Constraint for Photon Density Functionals}
\author{Xiaoyu Zhang}
\email{zhangxiaoyu@connect.hku.hk}
\affiliation{\normalfont Department of Chemistry, The University of Hong Kong, Pokfulam Road, Hong Kong SAR, China}
\begin{abstract}
When the coupling to a quantized field becomes weak, the leading term of the photon exchange--correlation action depends on the fluctuations of the interacting electronic system, not those of a density-matched noninteracting reference. We establish this requirement by proposing an adiabatic connection formulation of electron-photon coupling at fixed electronic density and controlling the density derivatives. Approximating the response in the integral yields a ring functional in terms of the Kohn--Sham response and electronic kernel. Exact model tests verify fourth-order rather than second-order errors. In a two-electron Hubbard dimer, the functional correctly predicts a correlation-induced polarization reversal, while two Kohn--Sham photon prescriptions don't. In an interacting quantum-dot resonator model with electron injection and escape, the functional recovers resonant photon-assisted transport missed by two Kohn--Sham photon prescriptions.
\end{abstract}
\maketitle

Electronic correlations can qualitatively alter electronic fluctuations and responses, with important applications. In an Anderson--Holstein model of a metal-coupled impurity, electronic correlations produce friction peaks absent from mean-field theory~\cite{Dou2017}. Noise calculations distinguish correlated-junction approximations more sharply than average currents~\cite{Miwa2017}. Coupled with photons, voltage-biased quantum-dot resonators provide a tunable simulator of molecular vibrational pumping, with energy transfer controlled by local electronic population inversion~\cite{Hermansen2024}. These examples motivate a precise question: which electronic fluctuations must a density functional retain when the coupling to a quantized field becomes weak?

Quantum-electrodynamical density-functional theory (QEDFT) supplies coupled electronic and photonic variables and auxiliary Kohn--Sham (KS) systems~\cite{Tokatly2013,Ruggenthaler2014,Farzanehpour2014}, with response formulations for their excitations~\cite{Flick2019,Yang2021}. For contacted systems, nonequilibrium Green-function methods describe transient currents and electronic correlations~\cite{Jauho1994,Myohanen2009}; time-dependent density-functional theory (TDDFT) provides open-boundary orbitals~\cite{Stefanucci2004,Kurth2005} and reduced-device density matrices~\cite{Zheng2007,Zheng2010}. Density--potential assumptions have been studied for open devices~\cite{Zheng2011} and fixed-point TDDFT constructions~\cite{Ruggenthaler2011}. Optical-response formulations connect transport and spectroscopy~\cite{Galperin2008,Gao2016}. Dissipative density-functional theories use stochastic or unitary auxiliary dynamics~\cite{DiVentra2007,YuenZhou2010}; their linear-response formulation distinguishes physical linewidths from bare KS linewidths through a bath-dependent kernel~\cite{Tempel2011}. Continuous photon spectra~\cite{Wang2021,Kudlis2022} and macroscopic electrodynamics~\cite{Svendsen2024} bring quantum optical environments into QEDFT.

Local photon potentials were constructed by extending time-dependent optimized effective potential (OEP) theory~\cite{Ullrich1995,vanLeeuwen1996} to photon-mediated interactions~\cite{Pellegrini2015,Flick2018}. Contour actions~\cite{vanLeeuwen1998}, conserving approximations~\cite{vonBarth2005,Tokatly2018}, and real-time OEP algorithms~\cite{Wijewardane2008,Liao2017} supply its formal and computational foundations. Other photon functionals use local polarizabilities~\cite{Flick2022,Mejia2025}, random-phase approximation (RPA)~\cite{Novokreschenov2023}, many-body dispersion~\cite{Tasci2025}, force-based photon-free constructions~\cite{Schafer2021,Lu2024,Ahmadabadi2026}, or quantum Monte Carlo~\cite{Weber2025}; model-functional bounds provide additional exact information~\cite{Bakkestuen2025}. Electronic screening also enters electron--photon Hedin equations~\cite{Trevisanutto2015}, electron--boson dynamics~\cite{deMelo2016,Karlsson2021}, and QED ring coupled cluster~\cite{DePrince2026}. Here we derive an adiabatic connection formulation on electron-photon coupling, use its lower endpoint to constrain the action and its derivatives, and integrate an explicit response approximation to obtain a photon functional.

In order to describe a device region that is coupled to electronic contacts and a quantized photon field, we start with the length--gauge Pauli--Fierz dipole Hamiltonian (cf. Eq.~(13) of Ref.~\cite{Tokatly2013}), with zero photon drive. Dropping the zero-point energy and replacing the electronic Hamiltonian, we have
\begin{equation}
 \hat H^{(s)}[v]=\hat H_\e[v]+\sum_\alpha\bigl[
 \omega_\alpha\hat a_\alpha^\dagger\hat a_\alpha
 +s\hat m_\alpha\hat F_\alpha+\tfrac{s^2\Lambda_\alpha}{2}\hat m_\alpha^2\bigr].
 \label{eq:H}
\end{equation}
For a device $C$ connected to leads $L_\ell$, the electronic region is $\Omega=C\cup\bigcup_\ell L_\ell$. We write $\hat H_\e[v]=\hat H_\e[0]+\int_\Omega\dd\bm r\,v(\bm r,t)\hat n(\bm r)$, where $v$ is the scalar potential acting on the entire electronic system, and $\hat n$ is the spin-summed electron density operator.
Each photon mode $\alpha$ is characterized by its frequency $\omega_\alpha>0$, annihilation operator $\hat a_\alpha$, and light-matter coupling vector $\bm\lambda_\alpha$. We define $g_\alpha=\sqrt{\omega_\alpha/2}|\bm\lambda_\alpha|$ and $x_\alpha(\bm r)=\bm\lambda_\alpha\cdot\bm r/|\bm\lambda_\alpha|$. Assuming that photons couple to the device region $C$, $\hat m_\alpha=\int_C\dd\bm r\,x_\alpha(\bm r)\hat n(\bm r)$. We further define $\hat F_\alpha=g_\alpha(\hat a_\alpha+\hat a_\alpha^\dagger)$ and $\Lambda_\alpha=|\bm\lambda_\alpha|^2$. A dimensionless parameter $s$ multiplies every light-matter coupling vector.

Throughout, $\langle\hat O\rangle_{\hat\varrho}\equiv\Tr(\hat\varrho\hat O)$ with $\Tr\hat\varrho=1$.

For real-time dynamics, we assume that (i) the initial
electron--photon state is factorized,
$\hat\varrho_0=\hat\varrho_{\e0}\otimes\hat\varrho_{\gamma0}$
at $t_0$; (ii) the initial photon state
$\hat\varrho_{\gamma0}$ is Gaussian;
(iii) the initial photon field has zero mean,
$\langle\hat a_\alpha\rangle_{\hat\varrho_{\gamma0}}=0$ for all $\alpha$;
and (iv) the initial state is kept fixed along
the photon-coupling path.

Let $\cC$ run from $t_0$ to $t_f$ and back, with contour time $z$, ordering $T_\cC$, and contour delta $\delta_\cC$. In kernel arguments, $1,2,\ldots$ label space--time points: $1=(\bm r_1,z_1)$, $\dd1=\dd\bm r_1\dd z_1$, with $\bm r_1\in C$. All operators are in the Schr\"odinger picture; time labels on insertions under $T_\cC$ specify their contour positions. For $\hat H_{\gamma0}=\sum_\alpha\omega_\alpha\hat a_\alpha^\dagger\hat a_\alpha$, the free photon propagator is
\[
 D_{\alpha\beta}(z,z')=-\ii\Big\langle T_\cC\hat F_\alpha(z)\hat F_\beta(z')
 e^{-\ii\int_\cC\dd\bar z\,\hat H_{\gamma0}}\Big\rangle_{\hat\varrho_{\gamma0}}.
\]
We define a photon-mediated electron-electron interaction as
\begin{equation}
\begin{split}
 \bar U_\gamma(1,2)={}&\sum_{\alpha\beta}x_\alpha(\bm r_1)x_\beta(\bm r_2)\\
 &\times[\Lambda_\alpha\delta_{\alpha\beta}\delta_\cC+D_{\alpha\beta}](z_1,z_2).
\end{split}
 \label{eq:kernel}
\end{equation}

For auxiliary $\lambda\in[0,s^2]$, define $Z_\lambda[v]=\langle T_\cC e^{-\ii\int_\cC\hat H^{(\sqrt\lambda)}[v]\dd z}\rangle_{\hat\varrho_0}$ and $U_\gamma=s^2\bar U_\gamma$. Eliminating photons gives
\begin{equation}
\begin{split}
 Z_\lambda[v]={}&\Big\langle T_\cC\exp\Big[-\ii\int_\cC\dd z\,\hat H_\e[v]\\
 &-\frac{\ii\lambda}{2}\int\dd1\dd2\,
       \hat n(1)\bar U_\gamma(1,2)\hat n(2)\Big]\Big\rangle_{\hat\varrho_{\e0}}.
\end{split}
 \label{eq:influence}
\end{equation}
Products denote contour convolution, $(AB)(1,2)=\int\dd3\,A(1,3)B(3,2)$, and $\Tr_\cC A=\int\dd1\,A(1,1)$.

We denote $v|_C=v_C$ and $v|_{L_\ell}=v_{\rm ext}|_{L_\ell}+\eta_\ell(z)$ for the device region and lead region, respectively. Here $v_{\rm ext}$ is the physical external potential and $\eta=\{\eta_\ell\}$; $\eta=0$ restores the physical lead potentials. For a prescribed device density $n$, the Legendre transform gives
\begin{equation}
\begin{gathered}
 \Gamma_\lambda[n;\eta]=-\ii\ln Z_\lambda[v_{\lambda,C},\eta]
       +\int\dd1\,v_{\lambda,C}(1)n(1),\\
 n(1)=\left.\ii\frac{\delta\ln Z_\lambda[v_C,\eta]}{\delta v_C(1)}
       \right|_{v_C=v_{\lambda,C}}.
\end{gathered}
 \label{eq:legendre}
\end{equation}
Here, $v_{\lambda,c}$ is the device potential that reproduces $n$ at $\eta$ and $\lambda$.
Differentiation gives
\begin{equation}
 \left.\partial_\lambda\Gamma_\lambda\right|_{n,\eta}
 =-\frac12\int\dd1\dd2\,\bar U_\gamma(1,2)
       [n(1)n(2)+\ii\chi_\lambda(2,1)],
 \label{eq:cancel}
\end{equation}
where 
\begin{equation}
     \chi_\lambda(1,2)=\left.\ii\frac{\delta^2\ln Z_\lambda}
 {\delta v_C(1)\delta v_C(2)}\right|_{v_C=v_{\lambda,C}}.
\end{equation}
For $A_\gamma=\Gamma_0-\Gamma_{s^2}$ and the mean-field term $A_\MF=\frac12\int\dd1\dd2\,n(1)U_\gamma(1,2)n(2)$, we define
\begin{equation}
 A_\xc^\gamma\equiv A_\gamma-A_\MF
 =\frac{\ii}{2}\int_0^{s^2}\dd\lambda\,
       \Tr_\cC[\bar U_\gamma\chi_\lambda[n;\eta]].
 \label{eq:ac}
\end{equation}
Fix $t_f<\infty$ and the initial state. Let $n_0$ denote the device-density history at $s=0$ and
$\eta=0$. Assume that the following conditions hold uniformly
on an $s$-independent neighborhood of the history
$(n,\eta)=(n_0,0)$:

(i) For every history, $v_{\lambda,C}[n;\eta]$ satisfying Eq.~\eqref{eq:legendre} exists for $0\le\lambda\le\epsilon$. 

(ii) The following mixed derivatives exist, are continuous, commute, and are uniformly bounded on these histories:
\[
 \sup_{0\le\lambda\le\epsilon}
 \left\|\partial_\lambda
 \frac{\delta^{p+q}}{\delta n^p\delta\eta^q}
 \Tr_\cC(\bar U_\gamma\chi_\lambda[n;\eta])\right\|\le M,
 \qquad p+q\le2.
\]
Here $p,q$ are nonnegative integers, $\epsilon,M>0$ are independent of $s$.

For $0\le s^2\le\epsilon$, Eq.~\eqref{eq:ac} gives
\begin{equation}
\begin{split}
 A_\xc^\gamma-\frac{\ii s^2}{2}\Tr_\cC(\bar U_\gamma\chi_0)
 &=\frac{\ii}{2}\int_0^{s^2}\dd\lambda\,(s^2-\lambda)\\
 &\quad\times\partial_\lambda\Tr_\cC(\bar U_\gamma\chi_\lambda).
\end{split}
 \label{eq:remainder}
\end{equation}
Differentiating this integral yields, uniformly over the same histories,
\begin{equation}
\begin{gathered}
 \left\|\frac{\delta^{p+q}}{\delta n^p\delta\eta^q}
 \left[A_\xc^\gamma-\frac{\ii s^2}{2}\Tr_\cC(\bar U_\gamma\chi_0)\right]\right\|\\
 \le\frac M2\int_0^{s^2}(s^2-\lambda)\dd\lambda
 =\frac M4s^4,\qquad p+q\le2.
\end{gathered}
 \label{eq:constraint}
\end{equation}

\begin{figure*}[t]
 \centering\includegraphics[width=.86\textwidth]{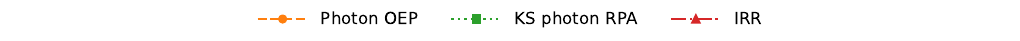}\\
 \includegraphics[width=.32\textwidth]{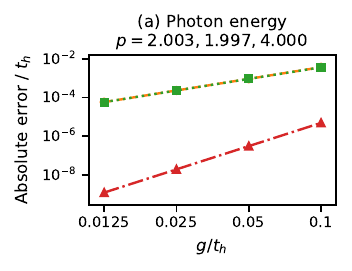}\hfill
 \includegraphics[width=.32\textwidth]{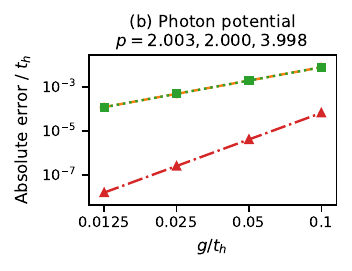}\hfill
 \includegraphics[width=.32\textwidth]{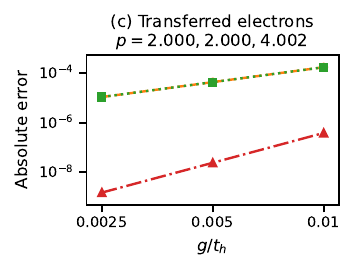}
 \caption{Weak-coupling accuracy: dimer photon energy $|E_\gamma^a-E_\gamma^{\rm ex}|/t_h$ (a), photon potential $|\partial_\rho(E_\gamma^a-E_\gamma^{\rm ex})|/t_h$ (b), and triangular-dot transferred electrons $\max_{0\le t\le T,\,\ell=L,R}|Q_\ell^a(t)-Q_\ell^{\rm ex}(t)|/e$ (c).}
 \label{fig:constraint}
\end{figure*}

Apply Eq.~\eqref{eq:legendre} to a noninteracting KS system representing the same density $n$. Its action $\Gamma_{\mathrm{s}}$ gives $v_{\mathrm{s}}=\delta\Gamma_{\mathrm{s}}/\delta n$ and $\chi_{\mathrm{s}}=(\delta^2\Gamma_{\mathrm{s}}/\delta n^2)^{-1}$. The electronic Hartree--xc kernel and reconstructed response are~\cite[Eqs.~(16),(17)]{vanLeeuwen1998}
\begin{equation}
\begin{gathered}
 f=\frac{\delta^2(\Gamma_{\mathrm{s}}-\Gamma_0)}{\delta n^2}
       =\chi_{\mathrm{s}}^{-1}-\chi_0^{-1},\\
 \chi=(\chi_{\mathrm{s}}^{-1}-f)^{-1}.
\end{gathered}
 \label{eq:electronic}
\end{equation}
We now start to introduce some approximations so that we can make calculations practical. Replace the integrand response in Eq.~\eqref{eq:ac} by $\chi_\lambda\approx(\chi^{-1}-\lambda\bar U_\gamma)^{-1}$, giving the interacting-response ring (IRR) functional
\begin{equation}
 A_\gamma^{\mathrm{IRR}}=A_\MF-\frac{\ii}{2}\Tr_\cC\ln(\mathbb I-U_\gamma\chi),
 \label{eq:ring}
\end{equation}
where $\mathbb I(1,2)=\delta(\bm r_1-\bm r_2)\delta_\cC(z_1,z_2)$. 

The contour extension of photon RPA~\cite{Novokreschenov2023} and the photon-exchange OEP action~\cite{Pellegrini2015} are, respectively,
\begin{equation}
\begin{aligned}
 A_\gamma^{\mathrm{RPA}}&=A_\MF-\frac{\ii}{2}\Tr_\cC\ln(\mathbb I-U_\gamma\chi_{\mathrm{s}}),\\
 A_\gamma^{\mathrm{OEP}}&=A_\MF+\frac{\ii}{2}\Tr_\cC(U_\gamma\chi_{\mathrm{s}}).
\end{aligned}
 \label{eq:comparators}
\end{equation}
Let $a\in\{\mathrm{IRR},\mathrm{RPA},\mathrm{OEP}\}$ index the three approximations. For $a=\mathrm{RPA},\mathrm{OEP}$,
\begin{equation}
 A_\gamma^a-A_\gamma=\frac{\ii s^2}{2}
 \Tr_\cC[\bar U_\gamma(\chi_{\mathrm{s}}-\chi_0)]+O(s^4).
 \label{eq:mismatch}
\end{equation}
In contrast, $\chi=\chi_0$ gives $A_\gamma^{\mathrm{IRR}}-A_\gamma=O(s^4)$. Similarly, the differences can be found in its derivatives.

For Eq.~\eqref{eq:ring}, variation at $\delta U_\gamma=0$ gives
\begin{equation}
\begin{split}
 \delta\chi&=\chi(\chi_{\mathrm{s}}^{-1}\delta\chi_{\mathrm{s}}\chi_{\mathrm{s}}^{-1}+\delta f)\chi,\\
 \delta(A_\gamma^{\mathrm{IRR}}-A_\MF)&=\frac{\ii}{2}\Tr_\cC[(\mathbb I-U_\gamma\chi)^{-1}U_\gamma\delta\chi].
\end{split}\label{eq:variation}
\end{equation}
We define $ u_\gamma=\frac{\delta A_\gamma^{\mathrm{IRR}}}{\delta n}$, and then we obtain
\begin{equation}
\begin{gathered}
 \chi_{\mathrm{s}}u_\gamma=
 \left.\frac{\delta A_\gamma^{\mathrm{IRR}}}{\delta v_{\mathrm{s}}}\right|_\eta,
 v_{\mathrm{s}}=v_{\rm ext}|_C+
 \frac{\delta(\Gamma_{\mathrm{s}}-\Gamma_0)}{\delta n}+u_\gamma.
\end{gathered}
 \label{eq:potential}
\end{equation}
The lead occupation and transferred charge, with $e>0$ the elementary-charge magnitude, are
\begin{equation}
\begin{gathered}
 N_\ell^{\mathrm{IRR}}=-\left.\frac{\delta(\Gamma_0-A_\gamma^{\mathrm{IRR}})}{\delta\eta_\ell}\right|_n,\\
 \frac{Q_L^{\mathrm{IRR}}(t)}e=N_L^{\mathrm{IRR}}(t_0)-N_L^{\mathrm{IRR}}(t),\\
 \frac{Q_R^{\mathrm{IRR}}(t)}e=N_R^{\mathrm{IRR}}(t)-N_R^{\mathrm{IRR}}(t_0),
\end{gathered}
 \label{eq:observable}
\end{equation}
for source $L$ and drain $R$. They satisfy
\[
 n^{\mathrm{IRR}}-n=O(s^4),\qquad (Q_\ell^{\mathrm{IRR}}-Q_\ell)/e=O(s^4).
\]

\begin{figure}[b]
 \centering\includegraphics[width=\columnwidth]{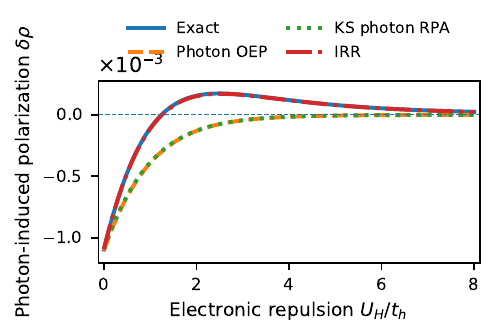}
 \caption{Correlation-induced polarization reversal, $\delta\rho=\rho(g)-\rho(0)$. }\label{fig:GS}
\end{figure}

\begin{figure*}[t]
 \centering\includegraphics[width=.98\textwidth]{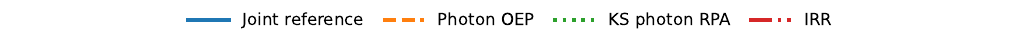}\\
 \includegraphics[width=.48\textwidth]{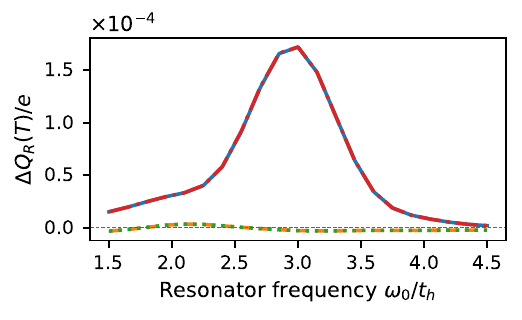}\hfill
 \includegraphics[width=.48\textwidth]{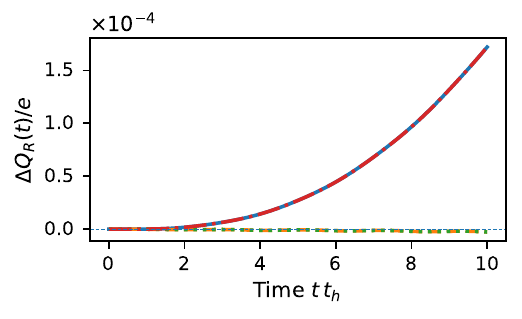}
 \caption{Resonant photon-assisted transport: extra drain electrons $\Delta Q_R(T)/e$ versus resonator frequency (left), and their buildup at resonance (right).}\label{fig:RT}
\end{figure*}

We revisit the two-electron Hubbard dimer~\cite{Carrascal2018} and the interacting triangular-dot resonator~\cite{Hermansen2024}. All three functionals use the same exact photon-free electronic input and the same physical parameters as the independent electron--photon reference. We set $\hbar=t_h=1$ and $g=sg_1$.

For the dimer~\cite[Eq.~(8)]{Carrascal2018},
\begin{equation}
 \hat H_\e[v]=-t_h\sum_\sigma(\hat c_{1\sigma}^\dagger\hat c_{2\sigma}+\mathrm{H.c.})
 +U_H\sum_i\hat n_{i\uparrow}\hat n_{i\downarrow}
 +\frac b2(\hat n_1-\hat n_2).
 \label{eq:dimer}
\end{equation}
Here $\hat c_{i\sigma}$ annihilates an electron on site $i=1,2$ with spin $\sigma\in\{\uparrow,\downarrow\}$, $\hat n_{i\sigma}=\hat c_{i\sigma}^\dagger\hat c_{i\sigma}$, and $\hat n_i=\sum_\sigma\hat n_{i\sigma}$. We consider a single photon mode in Eq.~\eqref{eq:H} and it has $C=\Omega$, $\hat m_1=\hat n_1-\hat n_2$, $\omega_1=\omega$. The density is $\rho=(n_1-n_2)/2$, with $n_i=\langle\hat n_i\rangle_{\hat\varrho}$. 

Exact diagonalization gives $E_g(b,g)=\min\operatorname{spec}\hat H^{(s)}[v]$. Density inversion fixes $b_g(\rho)$ by $\partial_bE_g(b_g,g)=\rho$, with $b_0=b_g|_{g=0}$. The fixed-density photon energy is
\[
\begin{split}
 E_\gamma^{\rm ex}(\rho;g)={}&E_g(b_g,g)-b_g\rho\\
 &-E_g(b_0,0)+b_0\rho.
\end{split}
\]
The corresponding approximations from Eqs.~\eqref{eq:ring} and \eqref{eq:comparators} are
\begin{equation}
\begin{gathered}
 E_\gamma^a(\rho;g)=\frac1{2\pi}\int_0^\infty\dd\xi\,
 \begin{cases}
  \ln(1+K_g\alpha_m),&a=\mathrm{IRR},\\
  \ln(1+K_g\alpha_{m,\mathrm{s}}),&a=\mathrm{RPA},\\
  K_g\alpha_{m,\mathrm{s}},&a=\mathrm{OEP},
 \end{cases}\\
 K_g(\xi)=\frac{2g^2\xi^2}{\omega(\xi^2+\omega^2)},\qquad \bm m^{\mathsf T}=(1,-1),\\
 \alpha_m=-\bm m^{\mathsf T}\chi_0\bm m,\qquad
 \alpha_{m,\mathrm{s}}=-\bm m^{\mathsf T}\chi_{\mathrm{s}}\bm m.
\end{gathered}
 \label{eq:GS}
\end{equation}
 The photon potential differences are
\[
\begin{aligned}
 u_{\gamma,1}^{\rm ex}-u_{\gamma,2}^{\rm ex}&=b_0-b_g,\\
 u_{\gamma,1}^a-u_{\gamma,2}^a&=\partial_\rho E_\gamma^a(\rho;g).
\end{aligned}
\]
At fixed $b$, the polarization in Fig.~\ref{fig:GS} follows from
\[
\begin{aligned}
 \rho^{\rm ex}(g)&=\partial_bE_g(b,g),\\
 \rho^a(g)&=\underset{\rho}{\operatorname{arg\,min}}\,
 [F_\e(\rho)+b\rho+E_\gamma^a(\rho;g)],
\end{aligned}
\]
where $F_\e(\rho)=E_g(b_0,0)-b_0\rho$ and $\delta\rho=\rho(g)-\rho(0)$. Figure~\ref{fig:constraint}(a),(b) fixes $U_H=2$, $\rho=0.25$, and $\omega=1$ at $g=0.0125,0.025,0.05,0.1$; Fig.~\ref{fig:GS} scans $0\le U_H\le8$ at $b=0.2$, $g=0.1$, and $\omega=1$.

For the triangular dots~\cite[Eqs.~(3),(5),(44)]{Hermansen2024},
\begin{equation}
\begin{split}
 \left.\hat H_\e[v_{\rm ext}]\right|_C={}&\epsilon_2\hat n_2
 +t_h(\hat c_1^\dagger\hat c_2+\hat c_3^\dagger\hat c_2+\mathrm{H.c.})\\
 &+t_d(\hat c_1^\dagger\hat c_3+\mathrm{H.c.})\\
 &+U\hat n_2(\hat n_1+\hat n_3)+\widetilde U\hat n_1\hat n_3.
\end{split}\label{eq:contact}
\end{equation}
Here $C=\{1,2,3\}$ contains the three dots, ordered from left to right; the external leads $L_L,L_R$ couple to sites $1,3$.
In $\left.\cdot\right|_C$, terms containing lead operators are omitted; photon terms remain.
The spin-polarized operators satisfy $\hat n_i=\hat c_i^\dagger\hat c_i$, with $v_{{\rm ext},i}=\epsilon_i$ and $\epsilon_1=\epsilon_3=0$.
The published capacitive coupling uses the form of Eq.~\eqref{eq:H} with $\hat m_1=\hat n_3$, $\omega_1=\omega_0$, and $\Lambda_1=0$.
We retain $t_d=0.5$, $U=20$, $\widetilde U=18$, $\Gamma=0.1$, $\kappa=0.005$, and the infinite-bias reservoirs and photon-loss dissipator.
We select the two-electron resonance at $\epsilon_2=-0.5$, $\omega_0=3$, reduce $g$ from $0.1$ to $0.01$, and calculate charge transfer after switching on $g$ at $t_0=0$.
Figure~\ref{fig:RT} scans $1.5\le\omega_0\le4.5$ and uses $\omega_0=3$ for its time trace; Fig.~\ref{fig:constraint}(c) uses $g=0.0025,0.005,0.01$ at resonance.
Both calculations use $T=10$ and $\Delta t=0.1$.

For $\mathcal D[\hat O]\hat\varrho=\hat O\hat\varrho\hat O^\dagger-\{\hat O^\dagger\hat O,\hat\varrho\}/2$, the exact reduced dot--photon state obeys
\begin{equation}
\begin{split}
 \dot{\hat\varrho}^{\rm ex}={}&-\ii[\left.\hat H^{(s)}[v_{\rm ext}]\right|_C,\hat\varrho^{\rm ex}]\\
 &+2\Gamma\{\mathcal D[\hat c_1^\dagger]+\mathcal D[\hat c_3]\}\hat\varrho^{\rm ex}
 +2\kappa\mathcal D[\hat a_1]\hat\varrho^{\rm ex}.
\end{split}\label{eq:lindblad}
\end{equation}

The corresponding electronic evolution is
\begin{equation}
\begin{gathered}
 \dot{\hat\varrho}_\e^a=-\ii\left[\left.\hat H_\e[v_{\rm ext}]\right|_C
       +\sum_{i\in C}u_{\gamma,i}^a\hat n_i,\hat\varrho_\e^a\right]\\
 \hspace{12mm}+2\Gamma\{\mathcal D[\hat c_1^\dagger]+\mathcal D[\hat c_3]\}\hat\varrho_\e^a,\\
 u_{\gamma,i}^a(t)=\left.\frac{\delta A_\gamma^a}{\delta n_i(t)}\right|_{n=n^a,\eta=0},\\
 n_i^{\rm ex}(t)=\langle\hat n_i\rangle_{\hat\varrho^{\rm ex}(t)},\qquad
 n_i^a(t)=\langle\hat n_i\rangle_{\hat\varrho_\e^a(t)}.
\end{gathered}\label{eq:functional_dynamics}
\end{equation}

We define the charge transfer as~\cite[Eq.~(30)]{Hermansen2024}
\begin{equation}
\begin{gathered}
 I_L(t)/e=2\Gamma[1-n_1(t)],\qquad I_R(t)/e=2\Gamma n_3(t),\\
 Q_\ell(t;g)=\int_0^t I_\ell(t';g)\dd t',\qquad \ell=L,R,\\
 \Delta Q_R(t)=Q_R(t;g)-Q_R(t;0).
\end{gathered}\label{eq:photocharge}
\end{equation}
These charges determine Figs.~\ref{fig:constraint}(c) and \ref{fig:RT}. 

Figure~\ref{fig:constraint} tests the photon energy, photon potential difference, and transferred electrons. The errors scale as $g^2$ for photon OEP and KS photon RPA but as $g^4$ for IRR.
In Fig.~\ref{fig:GS}, both KS photon approximations predict a negative polarization change where the exact change is positive. IRR reproduces the reversal with increasing electronic repulsion.
In Fig.~\ref{fig:RT}, IRR recovers both the resonant excess drain charge and its time dependence, whereas both KS photon approximations have almost no impact.

In conclusion, eq.~\ref{eq:constraint} ties the leading photon action and its derivatives to interacting electronic fluctuations in the weak coupling limit. Our functional, IRR, preserves this electronic response within a photon-ring sum. With exact electronic input, the two revisited models show fourth-order rather than second-order errors, the correct correlation-induced polarization reversal, and resonant photon-assisted charge transfer. Additional photon rings built from the bare KS response cannot replace the missing electronic correlations. This constraint should be useful for future photon functional constructions.

Codes for reproducing data are fully public at https://github.com/YuleZhang936/irr.

\label{maintextend}
\bibliography{main}
\end{document}